\documentclass[11pt,a4paper]{article}
\usepackage{jheppub}
\usepackage{epstopdf}

\usepackage{amssymb}
\usepackage{graphics,bm}
\usepackage{epsfig}
\usepackage{graphicx}
\usepackage{rotating}
\usepackage{color}

\newcommand{\beq}{\begin{equation}}
\newcommand{\eeq}{\end{equation}}
\newcommand{\bqa}{\begin{eqnarray}}
\newcommand{\eqa}{\end{eqnarray}}

\def\sumint{\hbox{$\sum$}\!\!\!\!\!\!\int}
\def\square{\vcenter{\vbox{\hrule height.4pt
          \hbox{\vrule width.4pt height4pt
          \kern4pt\vrule width.3pt}\hrule height.4pt}}}

\title{The chiral transition in a magnetic background: Finite density effects
and the functional renormalization group}

\author[\,a,b]{Jens O. Andersen}
\date{today}
\affiliation[a]{Department of Physics, 
Norwegian University of Science and Technology, 
H{\o}gskoleringen 5,
N-7491 Trondheim, Norway}
\emailAdd{andersen@tf.phys.ntnu.no}
\author[b]{Anders Tranberg}
\emailAdd{anders.tranberg@nbi.dk}
\affiliation[\,b]{
Niels Bohr International Academy, Niels Bohr Institute and Discovery Center,
Blegdamsvej 17, DK-2100 Copenhagen, Denmark}

\abstract{
We compute the phase diagram of the quark-meson model at finite temperature, 
finite baryon chemical potential $\mu_B=3\mu$ and constant 
external magnetic field $B$, 
using the functional renormalization group. 
Our results show that the critical temperature increases as a function of $B$ 
at $\mu=0$, but for values $\mu$ larger than about $210-225$ MeV, the 
opposite behavior is realized. As the magnetic field increases, the critical 
point $(T^*,\mu^*)$ moves from large $\mu$, small $T$ towards small 
$\mu$, larger $T$ in the $\mu$--$T$ phase diagram.
}

\begin{document}

\maketitle

%%%%%%%%%%%%%%%%%%%%%%%%%%%
\section{Introduction}
\label{sec:introduction}
%%%%%%%%%%%%%%%%%%%%%%%%%%%%

The phase structure of QCD as a function of temperature $T$ and baryon 
chemical $\mu_B$ has received a lot of attention.
%over the past decades
This interest has been spurred by the ongoing heavy-ion collision experiments 
at RHIC (Brookhaven)  and the LHC (CERN) and the search for the location of 
the critical endpoint where the curve of first-order chiral transitions 
terminates. Moreover, it turns out that the
phase diagram has a much richer structure than was anticipated.
In particular, there are a number of color superconducting phases at large 
baryon chemical potential and low temperature (see for
example the reviews~\cite{schmitt,fukushimi} for a thorough discussion).
%More recently, there has been quite some interest in the behavior of
%strongly interacting matter in external electromagnetic fields. The motivation
%to study the properties of QCD in external fields
%comes for example from astrophysics and heavy-ion collisions. 
More recently, there has been quite some interest in the behavior of
strongly interacting matter in external electromagnetic fields, motivated by 
astrophysical phenomena as well as heavy-ion collisions.
Large magnetic fields exist inside ordinary neutron stars as well as
magnetars~\cite{neutron}. 
Similarly, strong, but short-lived  magnetic fields, up to $B\sim 10^{19}$ 
Gauss or $|qB|\sim 6m_{\pi}^2$, may be generated during
noncentral heavy-ion collisions at RHIC and LHC~\cite{harmen,tonev,bzdak}.
%So far, most of the work on the thermodynamics of strongly interacting matter 
%in an external magnetic field has been done
%at vanishing baryon chemical potential and in this context
In the context of strongly interacting matter in an external magnetic
field, 
at least three important questions have arisen: Does the order of the
transition change with the strength of the magnetic field $B$?; how does the 
critical temperature $T_c$ 
depend on the 
strength of the magnetic field $B$?; and is there a splitting between 
the chiral transition and the deconfinement
transition beyond some finite value of the $B$-field? We will address the 
first two of these questions in the present work.

In Ref.~\cite{chiralB} chiral perturbation theory at leading order was used to 
study the chiral transition as a function of magnetic field.
Comparing the pressure of a hot pion gas  with that 
of a noninteracting quark-gluon
plasma with a subtracted vacuum energy term 
due to a nonzero
gluon condensate, %$\langle g^2G_{\mu\nu}G^{\mu\nu}\rangle$, 
they found a first-order
transition for weak magnetic fields.
The line of first-order transitions ends at a critical 
point $(\sqrt{|qB|},T)=(600,104)$ MeV, where $|q|$ is the pion electric charge.
For larger values of $|qB|$, the transition is a crossover. 

The authors of  Ref.~\cite{fragapol} use the Polyakov loop extended 
quark-meson (QM) model
to study the phase transition. The bosons are treated at tree level while the 
fermions are treated at the one-loop level, except that 
the vacuum fluctuations at zero magnetic field are neglected. The authors
report that they find a first-order
transition at the physical point if they keep the remaining $B$-dependent
vacuum fluctuations and a crossover if they are ignored.
The mean-field calculations of Ref.~\cite{rashid} 
as well as the functional renormalization group calculation of
Ref.~\cite{skokov} using the QM model
suggest that the transition remains a crossover. 
The conflicting result between Ref.~\cite{fragapol} and 
Refs.~\cite{rashid,skokov} could be due to a different treatment of
the vacuum fluctuations. In Refs.~\cite{friman,kylling}
the effects of the vacuum fluctuations on the chiral transition 
were studied for $B=0$, and it was shown that 
both the critical temperature $T_c$ and the 
order of transition depend on whether they are included or not.

Most model calculations conclude that the critical temperature is an 
increasing function
of the magnetic field $B$. This includes both 
mean-field type 
calculations~\cite{fraga1,fragapol,kenji,sado,pnjlgat,pnjlkas,duarte,pinto,jensoa}
involving (Polyakov-loop extended) Nambu-Jona-Lasinio (NJL)
and quark-mesons models as well
as chiral perturbation theory,
and beyond mean field
using functional renormalization group
techniques~\cite{skokov}. 
These results are in line with the lattice calculations
of Refs.~\cite{sanf,negro}, however, these were carried out for
bare quark masses that correspond to very large  pion masses in the
200-480 MeV range. 
The very recent simulations of Bali {\it et al} \cite{budaleik,gunnar}
for a pion mass of 140 MeV as well 
the MIT-bag calculations of Ref.~\cite{bag} suggest that the
critical temperature decreases with the strength of the magnetic field.

In order to address the issue of whether there is a splitting between
the chiral and deconfinement transitions, one must go beyond
the NJL or QM models since these 
%models 
have nothing to say about the latter.
This can be done by introducing an effective 
phenomenological potential for a constant temporal gluon field and expressing 
it in
terms of the thermal expectation value of the trace of the 
Polyakov loop~\cite{fukushima}. The conclusion of Ref.~\cite{fragapol}
is that the deconfinement transition temperature decreases with
the strength of the magnetic field, while the chiral transition temperature 
increases. Similar conclusions were reached in~\cite{skokov}, where a
functional renormalization-group approach was used together with
the P(QM) model to include mesonic fluctuations in the calculations.

In the present paper, we continue to explore the properties of QCD in
an external magnetic field using the quark-meson model. 
The work is a continuation of Ref.~\cite{rashid} in which the phase diagram
in the $\mu$--$T$ plane was mapped out for strong fields, 
$|qB|\sim 5m_{\pi}^2$.
The calculations were carried out in the mean-field approximation
and the large-$N_c$ limit where the bosonic modes are treated at tree level.
Moreover, the influence of the fermionic vacuum fluctuations on the
transition was studied. In the chiral limit, it was found that 
the transition is first order in the entire $\mu-T$ plane if 
vacuum fluctuations are not included and second order if they are included.
Here we are using the functional renormalization group (FRG)~\cite{wetterich}
to investigate the behavior of strongly interacting matter
in a magnetic background.
The functional
renormalization group is a non-perturbative, self-consistent resummation, 
with a wide range of 
applicability. See for example~\cite{rosten} for a recent review. 
%It is an alternative, but equivalent formulation of Wilson???s ideas 
%from the 1970s. 

The paper is organized as follows. In Sec.~\ref{sec:quarkmeson}, we briefly
discuss the quark-meson model and the functional renormalization group.
In Sec.~\ref{sec:numerical}, we present and discuss our numerical results 
while we
summarize in Sec.~\ref{sec:conclusion}. The renormalization group equation
for the effective potential is 
derived in Appendix~\ref{sec:flowequation}.

%%%%%%%%%%%%%%%%%%%%%%%%%%%%%%%%%%%%%%%%%%%%%%%%%%%%%%%%%%%%%%%%%%%%%%%%%%%
\section{Quark-meson model and the functional renormalization group}
\label{sec:quarkmeson}
%%%%%%%%%%%%%%%%%%%%%%%%%%%%%%%%%%%%%%%%%%%%%%%%%%%%%%%%%%%%%%%%%%%%%%%%

The $O(4)$-invariant 
Euclidean Lagrangian for the quark-meson model is
\bqa\nonumber
{\cal L}&=&
\bar{\psi}\bigg[
\gamma_{\mu}\partial_{\mu}-\mu\gamma_4
+g(\sigma-i\gamma_5{\boldsymbol \tau}\cdot{\boldsymbol \pi})\bigg]\psi
%{\cal L}_{\rm meson}+{\cal L}_{\rm quark}+{\cal L}_{\rm Yukawa}
%+{\cal L}_{\rm det}\;,
\label{lagra}
+{1\over2}\bigg[
(\partial_{\mu}\sigma)^2
+(\partial_{\mu}{\boldsymbol \pi})^2
\bigg]
+{1\over2}m^2\bigg[\sigma^2+{\boldsymbol \pi}^2\bigg]
\\ &&
+{\lambda\over24}\bigg[
\sigma^2+{\boldsymbol \pi}^2\bigg]^2
-h\sigma\;,
\eqa
%in terms of
%\bqa%\nonumber
%{\cal L}_{\rm meson}&=&
%%{1\over2}
%{\rm Tr}\left[
%\partial_{\mu}\Phi^{\dagger}\partial_{\mu}\Phi\right]
%+%{1\over2}
%m^2{\rm Tr}\left[\Phi^{\dagger}
%\Phi\right]
%+{\lambda\over3}{\rm Tr}\left[\Phi^{\dagger}\Phi\right]^2
%-{1\over2}h{\rm Tr}
%\left[\Phi+\Phi^{\dagger}\right]\;, \\
%{\cal L}_{\rm quark}&=&
%\bar{\psi}\left[
%\gamma_{\mu}\partial_{\mu}-\mu\gamma_4\right]\psi\;, \\
%{\cal L}_{\rm Yukawa}&=&
%g\bar{\psi}\left[\sigma-i\gamma_5{\boldsymbol \tau}\cdot{\boldsymbol\pi}
%\right]\psi\;,
%\\ 
%\label{det}
%{\cal L}_{\rm det}&=&
%c\det[\Phi+\Phi^{\dagger}]\;,
%\eqa
%where
%\bqa
%\Phi&=&{1\over2}\left(\sigma+
%{\boldsymbol \tau}\cdot{\boldsymbol\pi}\right)\;.
%\eqa
where $\sigma$ is the sigma field, ${\boldsymbol \pi}$ denotes the 
neutral and charged pions.
${\boldsymbol \tau}$ are the Pauli matrices,
$\mu=\mbox{$1\over2$}(\mu_u+\mu_d)$ 
is the quark chemical potential, in terms of
$\mu_u$ and $\mu_d$, the chemical potential for the $u$ and $d$ quarks,
respectively. The baryon chemical potential is given by $\mu_B=3\mu$. 
We set $\mu_u=\mu_d$ so that we are working at zero isospin
chemical potential, $\mu_I=\mbox{$1\over2$}(\mu_u-\mu_d)=0$.
The Euclidean $\gamma$ matrices
are given by $\gamma_j=i\gamma^j_M$, $\gamma_4=\gamma^0_M$,
and $\gamma_5=-\gamma^5_M$, where
the index $M$ denotes Minkowski space.
The fermion field is an isospin doublet
\bqa%\nonumber
\psi=
\left(\begin{array}{c}
u\\
d\\
\end{array}\right)\;.
\label{d0}
\eqa
If $h=0$, Eq.~(\ref{lagra}) is invariant under $O(4)$.
If $h\neq0$, chiral symmetry is explicitly broken, otherwise
it is spontaneously broken in the vacuum. Either way, the symmetry is
reduced to $O(3)$.
%~\footnote{\bf Note that in the presence of electromagnetism,
%the $SU(2)$ flavor symmetry is broken down to $U(1)_A$
%due to the different electric charges of the $u$ and $d$ quarks. 
%The expectation
%value of $\sigma$ then breaks this residual symmetry giving rise to
%a single Goldstone particle - the neutral pion. The charged pions
%are no}.
Note that this requires $m^2<0$ which is assumed in the
remainder of the paper. 

Chiral symmetry is broken in the vacuum by a nonzero
expectation value $\phi$ for the sigma field and we 
make the replacement
\bqa
\sigma\rightarrow\phi+\tilde{\sigma}\;,
\eqa
where $\tilde{\sigma}$ is a quantum fluctuating field.
The tree-level potential then becomes
\bqa
U_{\Lambda}&=&
{1\over2}m^2_{\Lambda}\phi^2
+{\lambda_{\Lambda}\over24}\phi^4
-h\phi\;.
\label{tree}
\eqa
Note that we have introduced a 
subscript $\Lambda$ on $U$,
$m^2$, and $\lambda$, where $\Lambda$ is the ultraviolet cutoff of the
theory (see section 3). This  
is a reminder that these are unrenormalized quantities~\footnote{
%The relation
%$h=f_{\pi}m_{\pi}^2$ implies that this quantity is known and that the term
%-h\phi$ can be added to the potential after we have determined 
%$\lambda_{\Lambda}$ and $m^2_{\Lambda}$. 
The symmetry breaking term
is equivalent to an external field that does not flow and 
therefore $h=h_{\Lambda}$.}.

%\section{Functional renormalization group}
We will follow Wetterich's~ implementation of the renormalization group
ideas based on the effective average action $\Gamma_k[\phi]$~\cite{wetterich}.
This action is a functional of a set of background fields that 
are denoted by $\phi$. $\Gamma_k[\phi]$ satisfies 
% a socalled flow equation which is 
an (integro-differential) flow equation in the variable $k$, to be specified below.
The subscript $k$ indicates that all the modes $p$ between the ultraviolet
cutoff $\Lambda$ of the theory and $k$ have been integrated out.
When $k=\Lambda$ no modes have been integrated out and $\Gamma_{\Lambda}$
equals the classical action $S$. On the other hand, when $k=0$, all the
momentum modes have been integrated out and $\Gamma_0$ equals the full
quantum effective action. The flow equation then describes the "flow" in the space of effective actions as a function of $k$.

In order to implement the renormalization 
group ideas, one introduces a regulator function $R_k(p)$.
The function $R_k(p)$
is 
large for $p < k$ and small for $p > k$ whenever $0 < k < \Lambda$, 
and $R_{\Lambda} (p) =\infty$. These properties ensure
that the modes below $k$ are heavy and decouple, and only the modes between k 
and the UV cutoff $\Lambda$ are light and integrated out.
The choice of regulator function
has been discussed in detail in the literature and some choices are
better than others due both to their analytical and stability properties 
.%\cite{kutofffunction}.
We briefly discuss our choice of regulator in Appendix A.

The flow equation for the effective action cannot be solved exactly
so one must make tractable and yet physically sound approximations.
%A systematic sequence of approximations is the derivative expansion.
The first approximation in a derivative expansion is the local-potential approximation
and in this case the flow equation for $\Gamma_k$ reduces to a
flow equation for an effective potential $U_k(\phi)$.
In the case of a constant magnetic field, 
the differential equation for $U_k$ reads~\cite{skokov}
\bqa\nonumber
\partial_k U_k
&=&
{k^4\over12\pi^2}
\left\{
{1\over\omega_{1,k}}\left[1+2n_B(\omega_{1,k})\right]
+{1\over\omega_{k,2}}\left[1+2n_B(\omega_{2,k})\right]
\right\}
\\ &&\nonumber
+k{|qB|\over2\pi^2}\sum_{m=0}^{\infty}{1\over\omega_{1,k}}
\sqrt{k^2-p^2_{\perp}(q,m,0)}\,\theta\left(k^2-p^2_{\perp}(q,m,0)\right)
\left[1+2n_B(\omega_{1,k})\right]
\\ && \nonumber
-{N_c\over2\pi^2}k\sum_{s,f,m=0}^{\infty}
{|q_fB|\over\omega_{q,k}}
\sqrt{k^2-p^2_{\perp}(q_f,m,s)}\,\theta\left(k^2-p^2_{\perp}(q_f,m,s)\right)
\left[1-n^+_F(\omega_{q,k})-n^-_F(\omega_{q,k})
\right]\;,
\\ &&
\label{flowu}
\eqa
where we have defined 
$\omega_{1,k}=\sqrt{k^2+U_k^{\prime}}\,$,
$\omega_{2,k}=\sqrt{k^2+U^{\prime}+2U_k^{\prime\prime}\rho}\,$,
$\omega_{q,k}=\sqrt{k^2+2g^2\rho}\,$,
$p^2_{\perp}(q,m,s)=(2m+1-s)|qB|\,$,
$n_B(x)=1/(e^{\beta x}-1)\,$, 
$n_F^{\pm}(x)=1/(e^{\beta(x\pm\mu)}+1)$, and $\rho={1\over2}\phi^2$.
%Here $\phi_k$ is the $k$-dependent minimum of the effective potential.
We briefly discuss the derivation of Eq.~(\ref{flowu}) in Appendix A.
If we neglect the bosonic fluctuations by ignoring the
first two lines of Eq.~(\ref{flowu}), we can solve the equation for
$U_k$ analytically. This yields the standard mean-field
result for the one-loop thermodynamic potential~\cite{rashid}.

At zero temperature, the Bose distribution function vanishes and the 
Fermi distribution function becomes a step function.
The flow equation~(\ref{flowu}) then reduces to
\bqa\nonumber
\partial_k U_k
&=&
{k^4\over12\pi^2}
\left\{
{1\over\omega_{1,k}}
+{1\over\omega_{2,k}}
\right\}
+k{|qB|\over2\pi^2}\sum_{m=0}^{\infty}
\sqrt{k^2-p^2_{\perp}(q,m,0)}\theta\left(k^2-p^2_{\perp}(q,m,0)\right)
\\ &&
-{N_c\over2\pi^2}k\sum_{s,f,m=0}^{\infty}{|q_fB|\over\omega_{q,k}}
\sqrt{k^2-p^2_{\perp}(q_f,m,s)}\theta\left(k^2-p^2_{\perp}(q_f,m,s)\right)
\left[1-\theta(\mu-\omega_{q,k})
\right]\;.
\label{flowvac}
\eqa
Furthermore, if we set $\mu=0$, the step function in~(\ref{flowvac})
vanishes and we obtain the flow equation in the vacuum.

%%%%%%%%%%%%%%%%%%%%%%%%%%%%%%%%%%%%%%
\section{Numerical results}
\label{sec:numerical}
%%%%%%%%%%%%%%%%%%%%%%%%%%%%%%%%

\begin{figure}
\begin{center}
\setlength{\unitlength}{1mm}
\includegraphics[width=10.0cm]{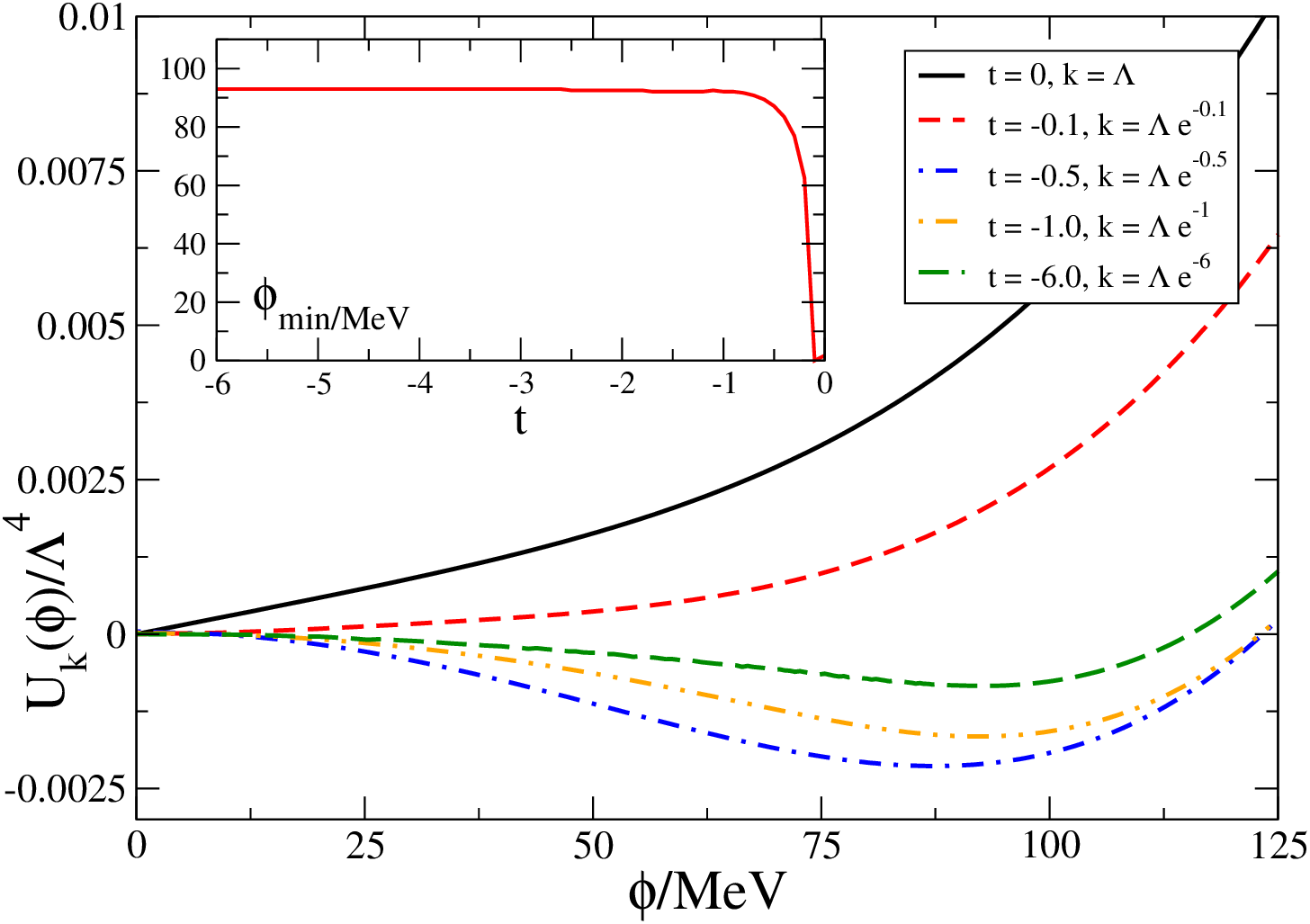}
\caption{Potential $U_k(\phi)$ 
for different values of $k$. Note that the curves do 
not evolve monotonically in $k$.
The inset shows the minimum $\phi_0=f_{\pi,k}$ as a function of 
$t=\log{k\over\Lambda}$.}
\label{integration}
\end{center}
\end{figure}

Solving the flow equation can in practice be done in two ways. One method is 
to make a polynomial expansion around the $k$-dependent
minimum and truncate the series, which leads to a set of coupled equations for 
the coefficients of the terms in this series. 
%This method fails near first-order transitions \cite{something}. 
The other possibility is to solve the flow equation
numerically for the potential $U_k(\phi)$ discretizing the field value $\phi$. 
We will do the latter, but discretizing instead in the quantity 
$\rho = \frac{\phi^2}{2}$
in the interval $\rho\in[0;8000]$ (MeV)$^2$, using $n=200$ grid 
points~\footnote{We checked that the results were insensitive to using half 
as many point, $n=100$.}. The probed interval corresponds to $\phi\in[0;126.5]$
MeV, which comfortably includes the minimum of the potential.

Earlier we pointed out that $h=h_{\Lambda}$ and it does not appear
in the flow equation~(\ref{flowu}). However, one can show that
$h=f_{\pi}m_{\pi}^2$ and so a nonzero  $h$ is introduced via the boundary
condition~(\ref{tree}) of the effective potential.
One must therefore tune the parameters $m_{\Lambda}$ and $\lambda_{\Lambda}$
separately for $h=0$ and $h\neq0$ to reproduce pion decay constant
$f_{\pi}$ and the pion mass $m_{\pi}$ in the vacuum correctly.
After tuning the parameters,
the tree level potential $U_{\Lambda}(\phi)$ is given by Eq.~(\ref{tree}), with 
$m_\Lambda^2=0$ throughout, and $\lambda_\Lambda=1.72$ 
for~\footnote{We will call this the ``$h>0$'' case or physical point.} 
$h=f_\pi m_\pi^2$, and $\lambda_\Lambda=2.125$ 
for~\footnote{$h=0$ is the chiral limit.} 
$h=0$. This ensures in both cases that at $T=\mu=B=0$, the minimum of the
potential is at $\phi_0=f_\pi=93$ MeV, and that the pion mass is correctly 
$m_{\pi}=140$ MeV for $h>0$, $h=0$, respectively. Also the $\sigma$ mass is in 
the allowed range of the broad resonance $m_{\sigma}=400-800$ MeV, and we 
specifically take 
$m_{\sigma}=450, 480$ MeV, again for $h>0$ and $h=0$. We use an ultraviolet 
cutoff of $\Lambda=500$ MeV as in Ref.~\cite{bernd}. 
We checked that changing the cutoff to 800 MeV results in a correction
of approximately 3\%
to $T_c$ at $\mu=0$, $B=0$, a precision we expect to persist throughout. The
bare coupling of course changes significantly, but the physical result
does not. 

We ignore the running of the Yukawa coupling and set 
$g_k=3.2258$ for all $k$. This yields a quark mass of 
$m_q=g\phi_0=gf_{\pi}=300$ MeV. The $k$-dependent masses 
$m^2_{\pi,k}$ and $m^2_{\sigma,k}$ are related to the $k$-dependent
minumum $\phi_{0,k}=f_{\pi,k}$ of the effective potential as follows
\bqa
\label{mpidef}
m_{\pi,k}^2=
{\partial^2U_{k}\over\partial\phi^2}\bigg|_{\phi=f_{\pi,k}},\qquad
m_{\sigma,k}^2=
m_{\pi,k}^2
+
\phi^2{\partial^2U_{k}\over\partial\phi^2}\bigg|_{\phi=f_{\pi,k}}\;.
\label{msigdef}
\eqa
The physical masses $m^2_{\pi}$ and $m^2_{\sigma}$ are given by evaluating
Eq.~(\ref{msigdef}) at $k=0$.
%In our calculations, we have used an ultraviolet cutoff $\Lambda=500$ MeV, 
%and this
Finally, when we integrate the flow equation, we use the dimensionless
variable $t\equiv\log{k\over\Lambda}$. Thus $k=\Lambda$ corresponds to
$t=0$ and $k=0$ corresponds to $t=-\infty$.

In Fig.~\ref{integration}, we show how the potential $U_k(\phi)$ develops for 
different
values of $k$. The black curve is the tree-level potential, $k=\Lambda$, and 
the green the full quantum effective 
potential $U_{0}(\phi)$. %~\footnote{We have checked that the potential
%has converged at $t=-6$.}. 
It is interesting to note that the potential at intermediate stages, here 
shown as red, blue, and
orange curves, do not evolve monotonically.
This shows that the bosonic and fermionic terms in the flow equation
dominate in different regions of integration.
It is important to point out that the exact effective potential
is both real and convex. It is an artefact of our truncation that 
the potential becomes nonconvex and that it develops an imaginary 
part~\cite{tetris}.

The inset shows the minimum $\phi_0=f_{\pi,k}$ as a function of the dimensionless
variable $t$. The minimum converges to $f_{\pi}$ around $t=-3$, but other 
features of the potential, such as the curvature in the potential minimum, 
continue to develop somewhat later. By $t=-6$ the potential has settled. 

\begin{figure}
\begin{center}
\setlength{\unitlength}{1mm}
\includegraphics[width=10.0cm]{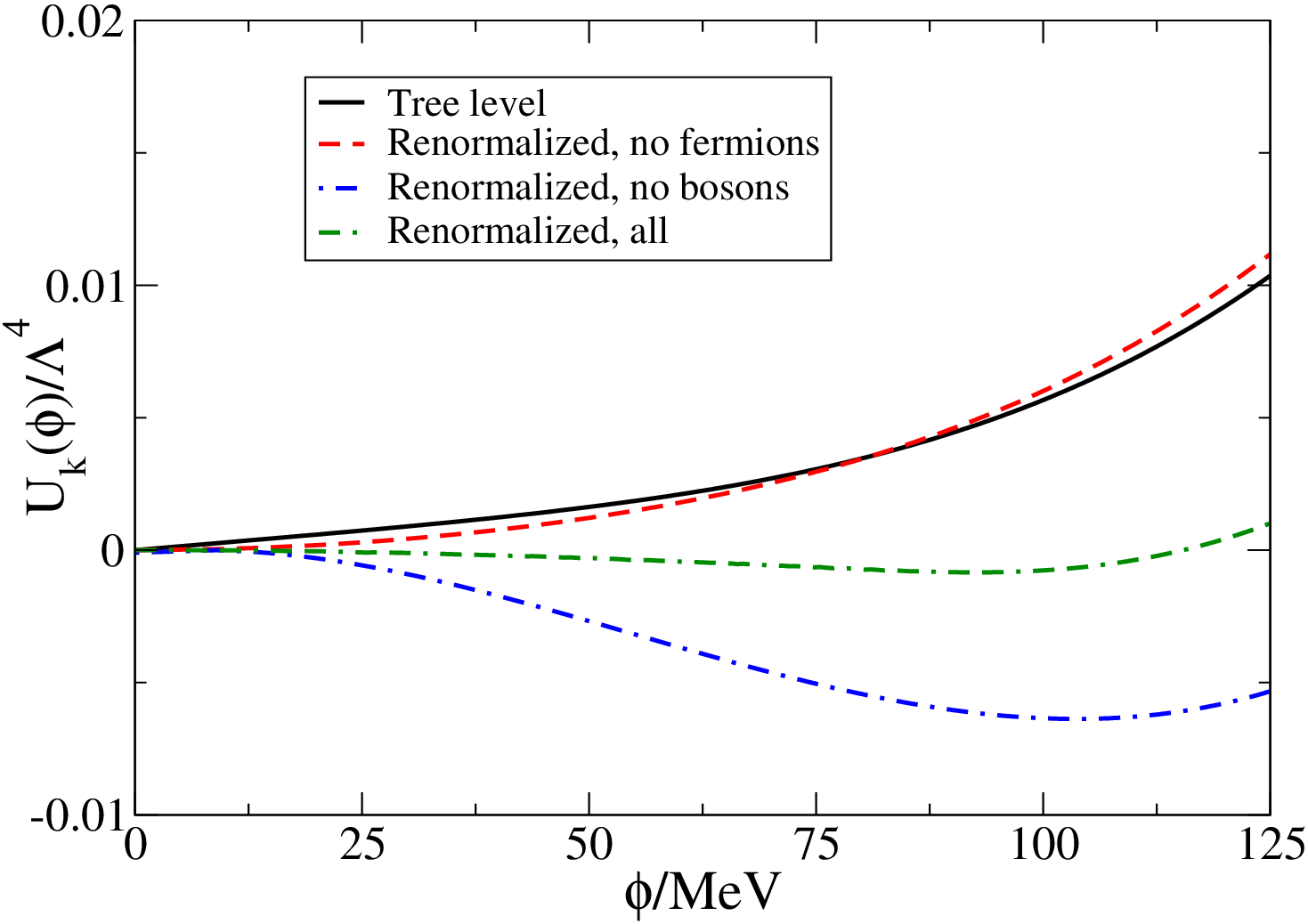}
\caption{Real part of the effective potential $U_k(\phi)$
for $\mu=T=B=0$. At tree-level $U_{\Lambda}(\phi)$ (black) and when fully 
renormalized $U_{k=0}(\phi)$ (green).
The blue and red curves are the effective potential where we have
ignored the effects of the bosonic quantum fluctuations and
fermionic vacuum fluctuations, respectively.}
\label{fluceff}
\end{center}
\end{figure}

In Fig.~\ref{fluceff}, we show various instances of the quantum effective 
potential
$U_k(\phi)$ for $\mu=T=B=0$, $h>0$, i.e. in the vacuum. The black line is the 
tree level potential $U_{k=\Lambda}(\phi)$ and the green line the fully 
renormalized effective potential $U_{k=0}(\phi)$. The red/blue lines are also 
fully renormalized, but ignoring the fermion/boson contribution to the 
fluctuations. We see that the two pull in opposite directions, 
enhancing/weakening symmetry breaking. 
The difference comes 
basically down to a different sign of the relevant terms in 
the flow equation~(\ref{flowvac}).
In some cases, the fermionic quantum fluctuations can destabilize the
effective potential altogether. Notice also that the green line is not simply 
the average of the red and the blue, as the flow equation performs non-linear 
resummations.

\begin{figure}
\begin{center}
\setlength{\unitlength}{1mm}
\includegraphics[width=10.0cm]{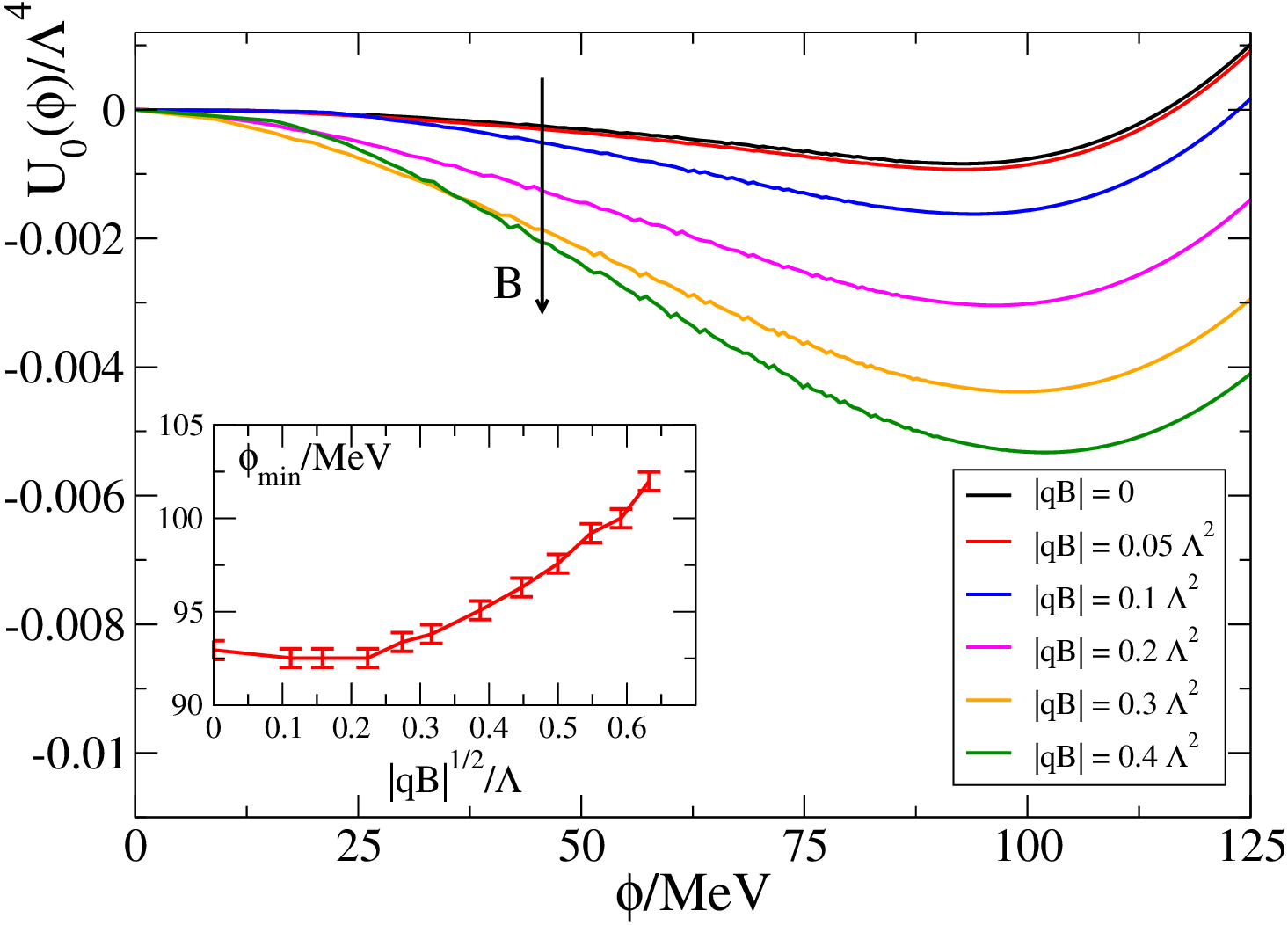}
\caption{Full quantum effective potential $U_0(\phi)$ for different values
of the magnetic field. 
The arrow indicates the direction of increasing
$B$.
Inset shows the minimum $\phi_0$ as a function
of the magnetic field $B$ with error bars.}
\label{vacb}
\end{center}
\end{figure}

Then in Fig.~\ref{vacb}, we show the effective potential
for different values of the magnetic field $B$ measured in units of
the ultraviolet cutoff $\Lambda^2$, at $T=\mu=0$, and here shown for $h>0$.
We notice that the effective potential becomes deeper and the minimum moves
to the right with increasing $B$. There is some roughness of the
effective potential for $\rho$ to the left of the minimum. The reason
is that the potential is complex to the left of the minimum
and numerical integration therefore becomes 
more noisy. We checked that this has no influence on the location of
the minimum of the potential.
The inset shows the minimum $\phi_0$
as a function of the strength of the magnetic field. 
The minimum is within discretization errors
an increasing, but non-linear, function of $B$. The error bars are meant to 
represent the resolution of the discretization in $\phi$. We found similar 
behaviour for $h=0$.
This result is in agreement with magnetic catalysis of dynamical symmetry
breaking (MCDS) which is the effect that chiral symmetry is broken
dynamically for any nonzero magnetic field when it is intact for $B=0$, and
more generally that a nonzero magnetic enhances symmetry breaking.
This effect is now well established in fermionic systems in an external
magnetic field, see for example 
Refs.~\cite{dcsm1,dcsm2,dcsm3,dcsm4,ebert22,dcsm5,dcsm6}.

\begin{figure}
\begin{center}
\setlength{\unitlength}{1mm}
\includegraphics[width=6.0cm]{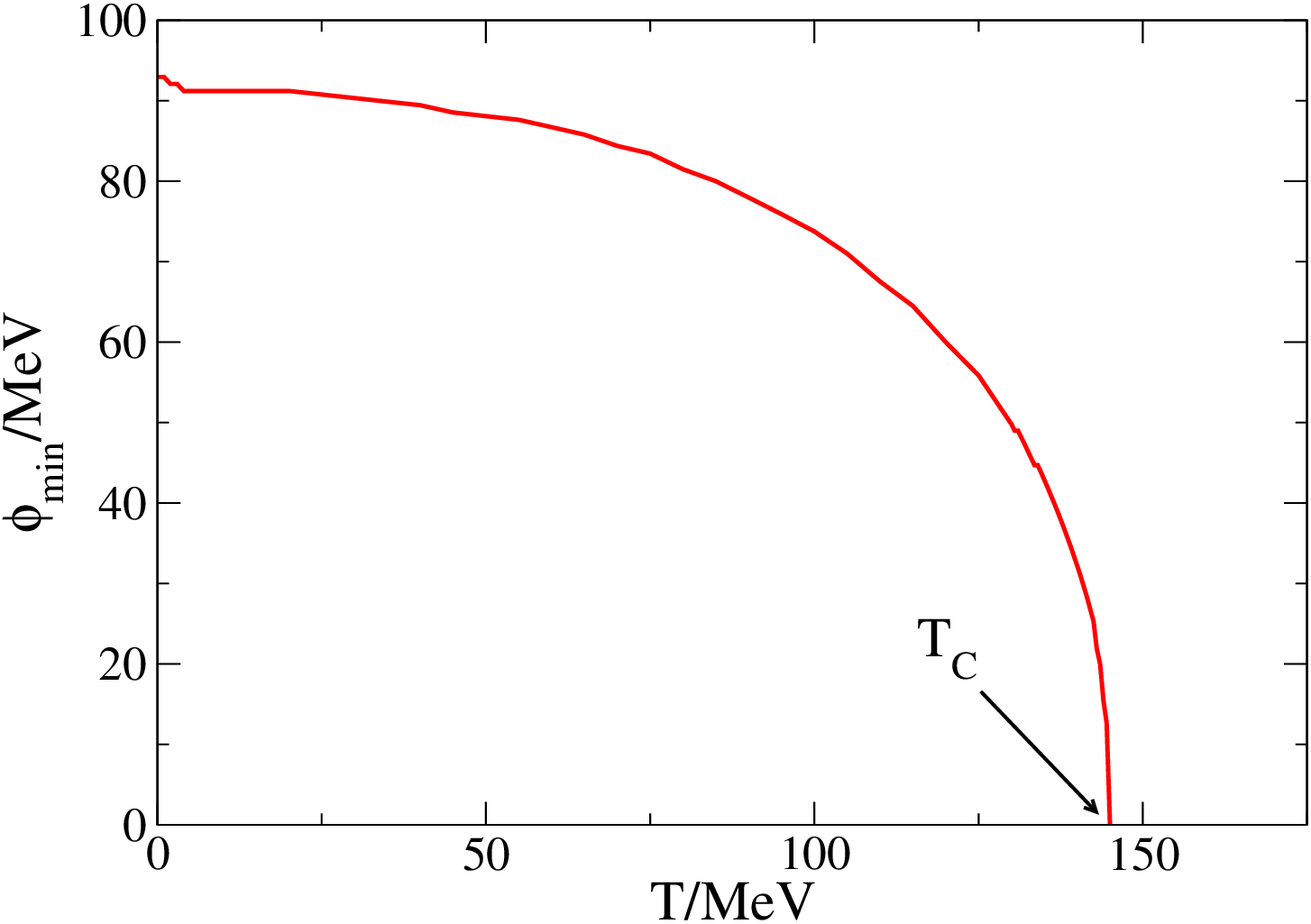}
\includegraphics[width=6.0cm]{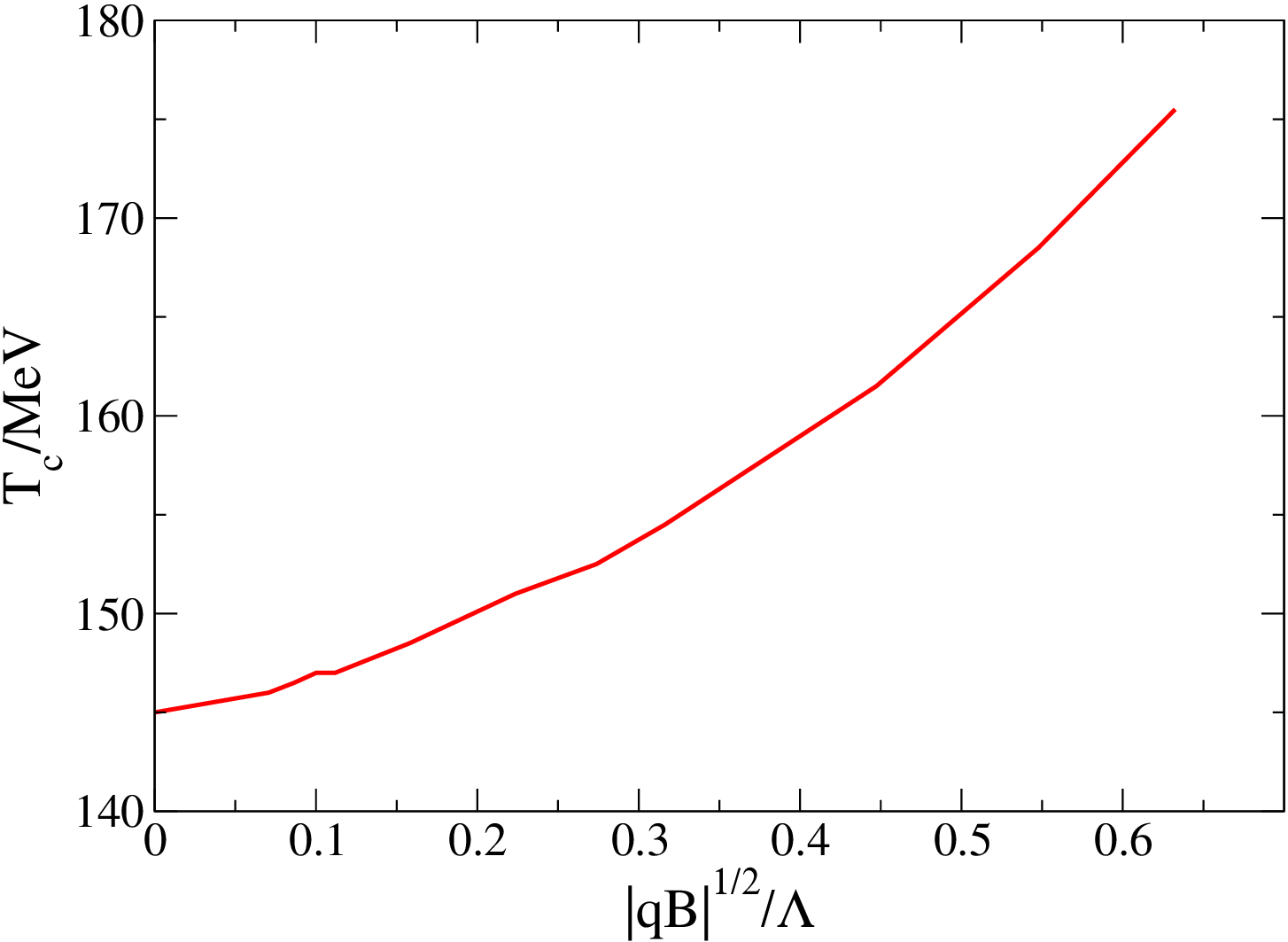}
\caption{Left: Minimum $\phi$ of the effective potential for $B=\mu=0$ at 
the physical point as a function of temperature $T$. Right: The critical 
temperature at $\mu=0$ as a function of $B$. $h>0$.}
\label{phioft}
\end{center}
\end{figure}

In the left-hand frame of Fig.~\ref{phioft}, we show the minimum of
the effective potential for $B=\mu=0$, $h>0$ as a function of temperature 
$T$, allowing for the determination of the critical temperature at this value 
of the chemical potential. The order parameter goes to zero in a continuous 
manner which shows that the transition is second order or a cross-over. The 
finite coarseness of the discretization does not allow us to firmly establish 
whether it is one or the other. 
%One may tentatively perform a fit of the form
%\bqa
%\phi_{\rm min}\propto |T-T_c|^\nu, 
%\eqa
%to find the critical exponent $\nu\simeq 0.4$. 
In the right-hand frame we show 
how this critical temperature depends on the magnetic field $B$. The 
dependence is monotonically, but non-linearly increasing. A similar behaviour 
was observed for $h=0$.

\begin{figure}
\begin{center}
\setlength{\unitlength}{1mm}
\includegraphics[width=10.0cm]{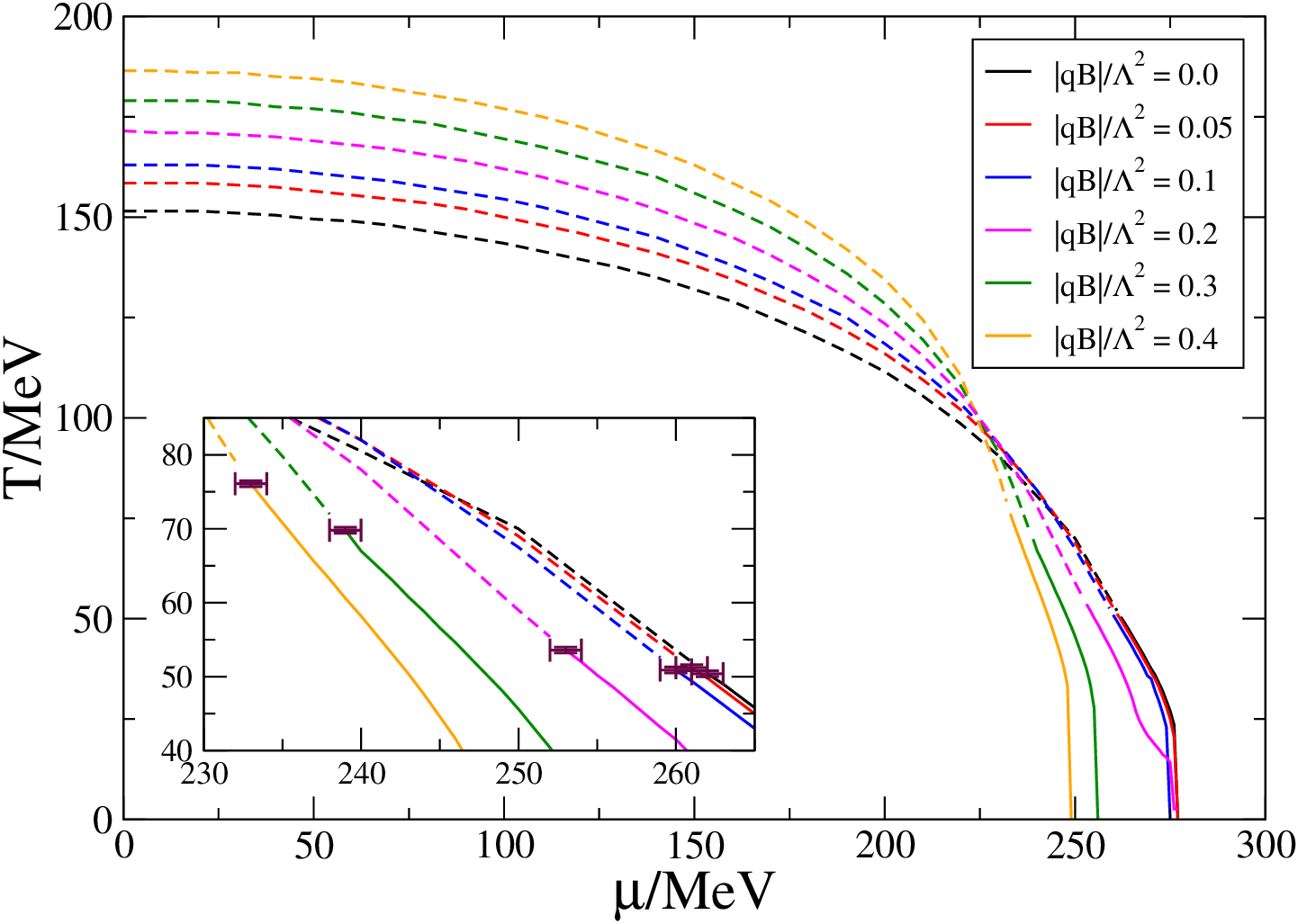}
\caption{Phase diagram in the $\mu$--$T$ plane in the chiral limit
for different values of the magnetic field.
The arrows indicate the direction of increasing $B$.
}
\label{phasen}
\end{center}
\end{figure}

Our first main result is then shown in Fig.~\ref{phasen}, displaying the phase 
diagram in the $\mu$--$T$
plane in the chiral limit, for a number of different value of the magnetic 
field $B$. The dashed lines show the part of the diagram where the transition 
is a cross-over, and the full lines where it is first order. At small chemical 
potential, increasing $B$ corresponds to increasing $T_c$. But interestingly, 
the lines cross as the large-$B$ curves dip down at much smaller $\mu_B$ than 
the small-$B$ curves. This means that at $\mu\geq225\,$MeV increasing $B$ 
at fixed $\mu$ corresponds to {\it decreasing} $T_c$. 

At very large chemical potential, the phase transition becomes first order at a
critical point (see inset). The crosses indicate the position of these 
critical points\footnote{Note that $T_c$ is the temperature at a given $\mu_B$ 
where there is a transition, but is not the temperature at the critical 
end point.} $(T^*,\mu^*)$. Since these appear in the range $\mu\geq225\,$MeV, 
the curves are ordered such that $T^*$ is an increasing function of $B$ and 
$\mu^*$ is a decreasing function of $B$. 

Additional features in the phase diagram have been reported in~\cite{bernd} 
at $B=0$, $h=0$ and large chemical potential, using slightly different 
parameters. In this region the authors found that the first-order transition
forks into two phase
transitions. The left transition line is always of first order while 
the chiral
restoration transition line around the splitting point is 
initially of first order
and turns into a second order transition for
lower temperatures.

Although we do see some possible additional structure appearing 
there, for instance in the $|qB|=0.2\Lambda^2$ curve, this region seems to be 
quite sensitive to the discretization of the numerics, and we feel unable to 
make any firm statements.

\begin{figure}
\begin{center}
\setlength{\unitlength}{1mm}
\includegraphics[width=10.0cm]{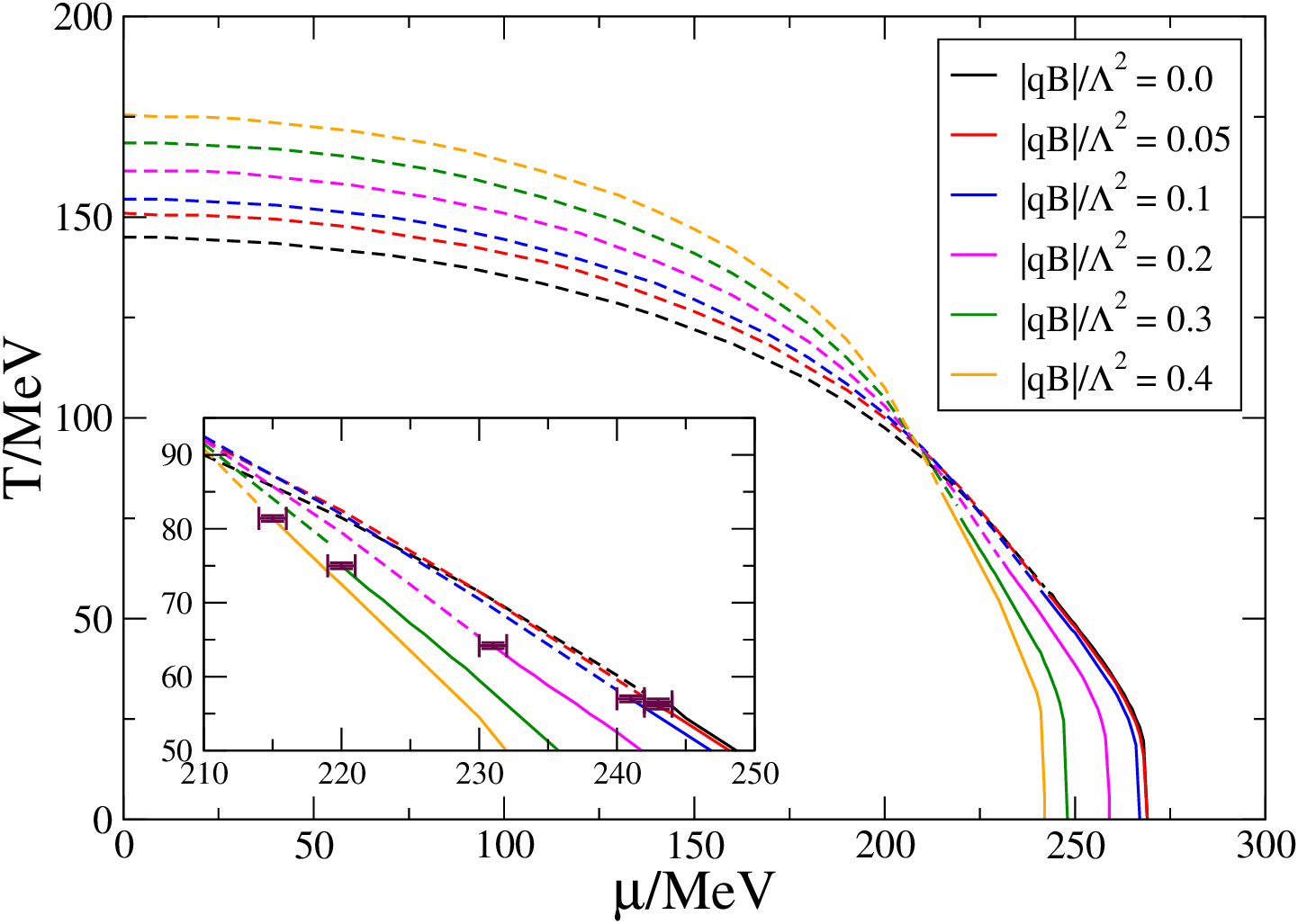}
\caption{Phase diagram in the $\mu$--$T$ plane at the physical point
for different values of
the magnetic field.
The arrows indicate the direction of increasing $B$.}
\label{phasen1}
\end{center}
\end{figure}

In Fig.~\ref{phasen1}, we show the phase diagram in the $\mu$--$T$ plane, but 
now
at the physical point $h>0$, again for different values of the magnetic field 
$B$. Again, the curves cross, now at a value of around $\mu=210$ MeV. In the 
inset, the crosses again indicate the position of the critical points. 

%%%%%%%%%%%%%%%%%%%%%%%%%%%%%%%%%%%%%%%%%
\section{Discussion and conclusion}
\label{sec:conclusion}
%%%%%%%%%%%%%%%%%%%%%%%%%%%%%%%%%%%%%%%%

In the present paper, we have mapped out the 
phase diagram of the quark-meson model in the presence of constant external
magnetic field using the functional renormalization group. 
For small values of the chemical potential, our study shows that 
the critical temperature is an increasing function of the magnetic field
$B$. However, for larger values of $\mu$ the curves for different $B$ cross 
in the $T-\mu$ plane and
so the critical temperature becomes a decreasing function of chemical potential
for large values
of $\mu$. This total inversion of the ordering of the curves takes place 
within a small range of chemical potentials of about $20$ MeV, centered around 
$225$ and $210$ MeV for $h=0$ and $h>0$, respectively. 
This behavior has also been seen in the NJL calculations 
of Refs.~\cite{inverse0,pinto} and the calculations of Ref.~\cite{inverse}
based on holography. The phenomenon was appropriately
dubbed ``inverse magnetic catalysis" (IMC).
Basically, there are two competing effects at work. 
Increasing the magnetic field $B$, effectively increases the 
particle-particle coupling and thereby enhances chiral symmetry 
breaking~\cite{dcsm2,dcsm3}. The chemical potential $\mu$ tends to 
split the particles and antiparticles and thus works against chiral
symmetry breaking. This is clearly demonstrated in the left half 
of Figs.~\ref{phasen} and~\ref{phasen1}; namely that for fixed $\mu$, the 
critical temperature increases with $B$, and for fixed $B$, the 
critical temperature decreases with $\mu$.
However, for larger values of $B$, the critical lines are crossing and
so the critical temperature decreases with $B$ (and still decreases with
$\mu$ as well). Thus the effects of increasing the magnetic field must
have changed qualitatively - namely that the cost of chiral symmetry
breaking exceeds the gain~\cite{inverse}.

Another main focus of our study was to pin down the critical endpoint, and 
its dependence on $B$. It turned out that the first-order line extends to 
higher temperatures for larger $B$; and because of the overall order-inversion 
of the transition curves the critical $\mu$ decreases with larger $B$. 
In particular, the largest magnetic field considered here corresponds to 
$|qB|\sim5m_\pi^2$, a value expected to be achievable in heavy-ion collisions. 
In this case, the critical endpoint should be expected at 
$(T^*,\mu^*)\simeq (81.4,215)$ MeV, rather than the $B=0$ values of 
$(56.6,243)$. In particular, this means that the first-order region can be
 reached at about 10-15 percent smaller baryon density.

In Ref.~\cite{buda}, the author have determined the phase diagram of
QCD in the $\mu$--$T$ plane for $B=0$ using lattice simulations
with physical quark masses and a Taylor-expansion technique
around $\mu=0$. The prediction for the critical line
is in reasonable agreement with the present results for small
values of the chemical potential but the curves diverge at around
$\mu=150$ MeV. Of course, for some (large) value of the baryon
chemical potential, the lattice results can no longer be trusted so the
discrepancy is not necessarily worrisome.

D'Elia {\it et al} have carried out lattice simulations in a constant
magnetic background at zero chemical potential~\cite{sanf}.
They explored various constituent quark masses corresponding to a pion mass
of $200-480$ MeV and different magnetic fields, up to $|qB|\sim20$ $m_{\pi}^2$
for the lightest quark masses.
For these values of the pion mass, 
they found that
there is a slight increase in the critical temperature $T_c$ for the
chiral transition. These results have been confirmed by
Bali {\it et al}~\cite{budaleik,gunnar}.
The same group has alo carried out 
lattice simulations for physical values of the pion mass, 
i.e. $m_{\pi}=140$ MeV. Their results which are extrapolated to the continuum
limit show that the 
critical temperature is a decreasing function of the magnetic field
\cite{budaleik,gunnar}. Hence the critical temperature 
for fixed $|qB|$ as a function
of the quark mass is nontrivial. 
Moreover, at $|qB|=1$ (GeV)$^2$ 
the transition is still a crossover, although somewhat stronger than at $B=0$
This is in stark contrast to most model calculations
that imply an increasing critical temperature as a functions of $B$.
This is irrespective of whether one goes beyond mean field or not.
Since this discrepancy is not understood, more work is needed to
resolve the problem.

%%%%%%%%%%%%%%%%%%%%%%%%%%%%%%%%%%%%%%%%%
\section*{Acknowledgments}
%%%%%%%%%%%%%%%%%%%%%%%%%%%%%%%%%%%%%%%%

J. O. A. would like to thank Gunnar Bali, 
Gergely Endr\H{o}di, and Falk Bruckmann
for useful discussions on their lattice simulations.
The authors would like to thank Andreas Schmitt, Toni Rebhan, Florian Preis, 
and Marcus Pinto for discussions on the inverse 
magnetic catalysis.
A. T. was supported by the Carlsberg Foundation.
J. O. A thanks the Niels Bohr International Academy and the Discovery 
Center for 
kind hospitality during the course of this work.

\appendix

%%%%%%%%%%%%%%%%%%%%%%%%%%%%%%%%%%
\section{Flow equation}
\label{sec:flowequation}
%%%%%%%%%%%%%%%%%%%%%%%%%%%%%%%%%

In this appendix, we briefly discuss the 
derivation of the flow equation~(\ref{flowu}).
The starting point is the exact flow equation
for the $k$-dependent effective action $\Gamma_k[\phi]$~\cite{wetterich} 
\bqa
\partial_k\Gamma_k[\phi]
&=&
{1\over2}{\rm Tr}\left[
\partial_kR_k(q)\left[
\Gamma_k^{(2)}+R_k(p)%_{q,-q}
\right]^{-1}
\right]\;,
\label{exactflow}
\eqa
where the superscript $n$ on $\Gamma_k[\phi]$ 
means the $n$th functional derivative of $\Gamma_k[\phi]$ and the trace is
over the spacetime momenta  $p$, 
and indices of the inverse propagator matrix. The function $R_k(p)$
is a regulator and is introduced in order to implement the renormalization 
group ideas: $R_k(p)$ is
large for $p < k$ and small for $p > k$ whenever $0 < k < \Lambda$, 
and $R_{\Lambda} (p) = \infty$. These properties ensure
that the modes below $k$ are heavy and decouple, and only the modes 
between $k$ and the UV cutoff
$\Lambda$ are light and integrated out. 
In order to proceed, we need to chose one regulator for the bosonic
fields and one for the fermionic fields. The choice of regulator has been 
discussed extensively in the literature
%~\cite{ball,jensmike,litim,canet,pawlow}
and in the present work
we will use~\cite{litim,stokic}
\bqa
\label{reg1}
R^B_k(p)&=&(k^2-{\bf p}^2)\theta(k^2-{\bf p}^2)\;,\\
R_k^F(p)&=&
\left(
\sqrt{{(p_0+i\mu)^2+k^2\over(p_0+i\mu)^2+{\bf p}^2}}-1\right)
(/\!\!\!p+i\mu\gamma^0)
\theta(k^2-{\bf p}^2)\;.
\label{reg2}
\eqa
This regulator is
very  convenient in practical
calculations since one can carry out the integral over 
three-momenta exactly. This turns the flow
equation into a partial differential equation as we show below.

The derivate expansion of the full quantum effective action reads
\bqa\nonumber
\Gamma[\phi]&=&
\int_0^{\beta}d\tau\int\,d^3x
\left\{
{1\over2}
Z_k^{(1)}\left[(\nabla\sigma)^2+(\nabla{\boldsymbol \pi})^2\right]
%\left[(\nabla\Phi)^{\dagger}\cdot(\nabla\Phi)\right]
+{1\over2}Z_k^{(2)}
\left[(\partial_0\sigma)^2+(\partial_0{\boldsymbol \pi})^2\right]
%\left[(\partial_0\Phi)^{\dagger}(\partial_0\Phi)\right]
\right.\\ &&\left.
+...+...U_k(\phi)
+Z^{(3)}_k\bar{\psi}[\gamma_{0}\partial_{0}-\gamma_4\mu]\psi
+Z^{(4)}_k\bar{\psi}\gamma_{i}\partial_{i}\psi+
g_k\bar{\psi}\left[\sigma-i\gamma_5{\boldsymbol \tau}\cdot{\boldsymbol\pi}
\right]\psi
+...\right\}\;,
\eqa
where $Z_k^{(i)}$ are wavefunction renormalization constants,
$U_k(\phi)$ is the scale-dependent effective potential, and
the ellipsis denote higher-order derivative terms that satisfy the
symmetries of the effective action. 
Note that the
$Z_k^{(i)}$ are in principle different since 
Lorentz invariance is broken due to both finite temperature and finite density.
In the local-potential approximation, one neglects the scale dependence
of $Z_k^{(i)}$ and retain their tree-level values $Z_k^{(i)}=1$.
The flow equation then reduces to a coupled equation for the effective
potential $U_k(\phi)$ and the Yukawa coupling $g_k$. If we furthermore ignore
the running of the latter, we obtain a tractable equation for 
the former. Ignoring the running of the Yukawa coupling, implies that 
the approximation does not correspond
to the local-potential approximation~\cite{bernd}.

The effective potential $U_k(\phi)$
is defined by evaluating the effective action 
$\Gamma_k[\phi]$ for space-time independent fields and dividing by the 
$VT$, where $V$ is the volume:
\bqa
U_k(\phi_{\rm uni})&=&{1\over\beta V}\Gamma_k[\phi_{\rm uni}]\;.
\label{udef}
\eqa
In the following, we neglect the subscript $\rm uni$.
Inserting Eq.~(\ref{udef}) into Eq.~(\ref{exactflow}), 
we obtain the flow equation for $U$. This yields
\bqa\nonumber
{\partial}_kU_k&=&
{1\over2}
\sumint_P{\rm Tr}
\left\{
\partial_kR^B_k(p)\left[
\left(p^2+R^B_k(p)\right)\delta_{ij}
+{\partial^2U_k\over\partial\Phi_i\partial\Phi_j}
\right]^{-1}
\right\}
\\ &&
+{1\over2}
\sumint_{\{P\}}{\rm Tr}\left\{
\partial_kR^F_k(p)
\left[
/\!\!\!\!P-\gamma_4\mu+g\phi+R_k^F(p)
\right]^{-1}\right\}
\;,
\label{matrix}
\eqa
where $\Phi=(\sigma,{\boldsymbol \pi})$, $i,j=1,2,3,4$, 
and
the trace is over internal indices only.
The symbols $\sumint_P$ and $\sumint_{\{P\}}$ 
are defined by
\bqa
\sumint_P&=&T
\hspace{-0.3cm}
\sum_{P_0=2\pi nT}\int{d^3p\over(2\pi)^3}\;,\\
\sumint_{\{P\}}&=&T
\hspace{-0.4cm}
\sum_{P_0=(2n+1)\pi T}\int{d^3p\over(2\pi)^3}\;.
\eqa

Using the chain rule, the matrix appearing in the first
term in Eq.~(\ref{matrix})
\bqa
\left(\begin{array}{cccc}
p^2+R_k^B(p)
+U_k^{\prime}+U_k^{\prime\prime}&0&0&0\\
&&&\\
0&p^2+R_k^B(p)+U_k^{\prime}&0&0\\
0&0&p^2+R_k^B(p)+U_k^{\prime}&0\\
0&0&0&p^2+R_k^B(p)+U_k^{\prime}\\
\end{array}\right)\;,
\label{matrix2}
\eqa
where %$F_k(p)=p^2+R_k^B(p)$ and 
$U_k^{\prime}={\partial U_k\over\partial\rho}$
with $\rho={1\over2}\phi^2$.
We notice that the matrix Eq.~(\ref{matrix2}) is simply the
inverse tree-level propagator if we make the substitution
${\cal V}_0\rightarrow U_k+R_k^B(p)$.
Integration over ${\bf p}$ gives
\bqa\nonumber
\partial_kU_k
&=&
{k^4\over12\pi^2}T\sum_{n=-\infty}^{\infty}
\left[
{3\over\omega_n^2+k^2+U^{\prime}_k}+
{1\over\omega_n^2+k^2+U^{\prime}_k+2\rho U^{\prime\prime}_k}
\right]\\
&&
-{N_cN_fk^4\over6\pi^2}T\sum_{n=-\infty}^{\infty}
\left[
{1\over(\omega_n+i\mu)^2+k^2+m_q^2}
{1\over(\omega_n-i\mu)^2+k^2+m_q^2}
\right]\;.
\eqa
%where we have changed variables, $t=\ln{k\over\Lambda}$.
Summing over the Matsubara frequencies, we obtain
\bqa\nonumber
\partial_kU_k
&=&
{k^4\over12\pi^2}
\left\{
{3\over\omega_{1,k}}\left[1+2n_B(\omega_{1,k})\right]
+{1\over\omega_{2,k}}\left[1+2n_B(\omega_{2,k})\right]
\right\}
%\\&&
\\ &&
-{N_cN_fk^4\over3\pi^2}
\left\{\left[
{1\over\omega_{q,k}}\left(1-n^+_F(\omega_{q,k})
-n^-_F(\omega_{q,k}\right)
\right]\right\}\;,
\label{flow0b}
\eqa
where $\omega_{1,k}=\sqrt{k^2+U_k^{\prime}}$, 
$\omega_{2,k}=\sqrt{k^2+U_k^{\prime}+\phi^2U_k}$, 
$\omega_{q,k}=\sqrt{k^2+g^2\phi^2}$, 
$N_B(x)=1/(e^{\beta x}-1)$.
and
$N_F^{\pm}(x)=1/(e^{\beta(x\pm\mu}+1)$.

We next briefly consider the flow equation in a constant magnetic field 
$B$~\cite{skokov}.
As we pointed out above, the matrices appearing on the right-hand side
of the flow equation are essentially the inverse tree-level
propagators. We can therefore find the flow equation
in a constant magnetic background since the tree-level propagators
are expressed in terms of the well-known solutions to the Klein-Gordon and
Dirac equations in constant $B$ field.
The spectra are given by $E^2=p_z^2+(2m+1)|qB|$ for bosons
and $E^2=p_z^2+(2m+1-s)|q_fB|$ for fermions, 
where $q_z$ is the $z$-component of 
the three-momentum,
$m$ is the $m$th Landau level
and $s=\pm1$ is the spin variable. 
We therefore make the substitutions
${\bf q}^2\rightarrow q_z^2+(2m+1)|qB|$ 
and $q_z^2+(2m+1-s)|q_fB|$ in the regulators~(\ref{reg1}) and~(\ref{reg2}).
In the case of the charged particles 
the sum-integral is replaced by a sum over Matsubara 
frequencies $P_0=2\pi nT$,
a sum over Landau levels $m$, and an integral over momenta $p_z$
in $d=1$
dimension:
\bqa
\sumint_P&\rightarrow&{|qB|T\over2\pi}
\hspace{-0.3cm}
\sum_{P_0=2n\pi T,}\sum_{m=0}^{\infty}
\int{dp_z\over2\pi}
\;,\\
\sumint_{\{P\}}&\rightarrow&{|qB|T\over2\pi}
\hspace{-0.3cm}
\sum_{P_0=(2n+1)\pi T,}\sum_{m=0}^{\infty}
\int{dp_z\over2\pi}\;.
\eqa
After integrating over $p_z$ and summing over Matsubara
frequencies, one obtains.
\bqa\nonumber
\partial_k U_k
&=&
{k^4\over12\pi^2}
\left[
{1\over\omega_{1,k}}\left(1+2n_B(\omega_{1,k}\right)
+{1\over\omega_{2,k}}\left(1+2n_B(\omega_{2,k}\right)
\right]
\\ && \nonumber
+{|qB|\over2\pi^2}\sum_{m=0}^{\infty}
{k\over\omega_{1,k}}
\sqrt{k^2-p^2_{\perp}(q,m,0)}\,\theta\left(k^2-p^2_{\perp}(q,m,0)\right)
\left[1+2n_B(\omega_{1,k})\right]
\\ &&\nonumber
-{N_c\over2\pi^2}\sum_{s,f,m=0}^{\infty}{|q_fB|k\over\omega_{q,k}}
\sqrt{k^2-p^2_{\perp}(q_f,m,s)}\,\theta\left(k^2-p^2_{\perp}(q_f,m,s)\right)
\left[1-n_F^+(\omega_{q_f,k})-n_F^-(\omega_{q_f,k}
)\right]\;,
\\ &&
\label{flowb}
\eqa
where we have defined $p^2_{\perp}(q,m,s)=(2m+1-s)|qB|$.
We close the Appendix by taking the 
limit $B\rightarrow0$ in Eq.~(\ref{flowb}). 
We change variable $p_{\perp}^2=2|qB|m$, which
yields $p_{\perp}dp_{\perp}=|qB|dm$. Replacing the sum by an integral, we
obtain
\bqa\nonumber
\partial_k U_k
&=&
{k^4\over12\pi^2}
\left[
{1\over\omega_{1,k}}\left(1+2n_B(\omega_{1,k}\right)
+{1\over\omega_{2,k}}\left(1+2n_B(\omega_{2,k})\right)
\right]
\\ && \nonumber
+{k\over2\omega_{1,k}\pi^2}\int_0^{\infty}dp_{\perp}p_{\perp}
\sqrt{k^2-p^2_{\perp}}\theta\left(k^2-p^2_{\perp}\right)
\left[1+2n_B(\omega_{1,k})\right]
\\ &&
-{N_cN_fk\over\omega_{q,k}\pi^2}\int_0^{\infty}dp_{\perp}p_{\perp}
\sqrt{k^2-p^2_{\perp}}\theta\left(k^2-p^2_{\perp}\right)
\left[1-n_F^+(\omega_{k})-n_F^-(\omega_{q,k}
)\right]\;.
\label{flowb2}
\eqa
Finally, integrating over $p_{\perp}$, the flow equation~(\ref{flowb2}) 
reduces to Eq.~(\ref{flow0b}).

%%%%%%%%%%%%%%%%%%%%%%%%%%%%%%%%%

\end{document}